\documentclass[useAMS,usenatbib,usegraphicx]{mn2e}

\usepackage{amsmath}

\title[]{BOUNDARY SHEAR ACCELERATION IN THE JET OF MKN501}
\author[S. Sahayanathan]{S. Sahayanathan\thanks{E-mail:
sunder@barc.gov.in}\\
Astrophysical Sciences Division, Bhabha Atomic Research Centre, Mumbai - 400085, India\\}
\begin{document}



\maketitle

\label{firstpage}

\begin{abstract}
The high resolution image of the jet of the BL Lac object MKN501 in radio, show
a limb-brightened feature. An explanation of this feature as an outcome of differential 
Doppler boosting of jet spine and jet boundary due to transverse velocity structure of the 
jet requires large viewing angle. However
this inference contradicts with the constraints derived from the high energy $\gamma$-ray 
studies unless the jets bends over a large angle immediately after the $\gamma$-ray zone
(close to the central engine). In this letter we propose an alternate explanation to the 
limb-brightened feature of MKN501 by considering the diffusion of electrons accelerated 
at the boundary shear layer into the jet medium and this consideration does not require 
large viewing angle. Also the observed difference in the spectral index at the jet boundary 
and jet spine can be understood within the frame work of shear acceleration.
\end{abstract}


\section{Introduction}

BL Lac objects are the extreme class of active galactic nuclei(AGN) with weak or no
emission lines and are categorized along with flat spectrum radio quasars(FSRQ) as
blazars. Their spectra cover a broad range of photon energies starting from radio to
gamma rays with a few of them detected in TeV energies by ground based Air Cerenkov
experiments(\cite{kraw04, katar01, samb00},\cite{costa02}). These sources are 
found to be strongly variable with flare time scales ranging from days to less than an hour
(\cite{gaidos96, coppi99, samb00, kraw00}). The short time variability 
and their detection at very high energies demand that the emission region should be moving 
down a jet at relativistic velocities close to the line of sight of the 
observer (\cite{ghisellini93, dondi95}). The strong polarization 
detected in radio/optical energies and the non-thermal photon spectra indicates the radio to x-ray 
spectra is due to synchrotron radiation from a non-thermal electron distribution cooling in a magnetic 
field. However the gamma ray emission from these sources is still not well understood.
Leptonic models explain the high emission as inverse Compton scattered synchrotron photons 
by the electron population responsible for the synchrotron process itself(SSC)
(\cite{maraschi92, bloom96, bottcher00}) where as in hadronic models it is due to the 
synchrotron proton emission and proton-photon interactions involving an external photon field 
(synchrotron proton blazar model(SPB))
(\cite{mannheim98, mucke03}). Under unification hypothesis of radio-loud AGN, BL Lac objects are 
considered to be aligned jet version of Fanaroff-Riley type I (FRI) radio galaxies (\cite{urrypado95}).

MKN501 is a nearby BL Lac object (z=0.034) and also the second extra galactic source 
detected in TeV photon energies by ground based Cherenkov Telescopes(\cite{quinn96}). It was later 
detected in MeV photon energies by the satellite based experiment EGRET (\cite{katoka99}). 
The radio images of MKN501 show a jet emerging from a bright nucleus (\cite{edward00,
giovanni99, aaron99, giroletti04}). 
The high resolution (milli arc second) radio images show a transverse jet structure with the 
edges being brighter than the central spine commonly referred as "limb-brightened" structure
(\cite{edward00, giovanni99, giroletti04}).
This feature is usually explained by the "spine-sheath" model where the velocity at the jet 
spine is larger compared to the velocity at the boundary. Such a radial stratification of 
velocity across the jet arises when jet moves through the ambient medium and the viscosity involved 
will cause a shear at the boundary. Three-dimensional hydrodynamic simulations of 
relativistic jets (\cite{aloy00}) and two-dimensional simulations of relativistic magnetized jets
(\cite{leismann05}) also supports the presence of jet velocity stratification due to its 
interaction with the ambient medium. The existence of velocity 
shear at the jet boundary was first suggested by \cite{owen89} to explain the morphology of M87 jet.
\cite{perlman99} later confirmed it through the polarisation studies of M87 jet. 
If the jet is misaligned towards 
the observer, it may happen for a proper combination of velocities we see a Doppler boosted
image of the boundary compared to the less boosted spine giving rise to a limb-brightened
structure(\cite{komissarov90,laing96}). 
A possible consequence of the velocity shear is the alignment of the magnetic field at the 
boundary parallel to 
the flow velocity due to stretching of the frozen-in field lines of the plasma(\cite{kahn83}). 
The polarisation angle observed at the jet boundary of MKN501 is perpendicular to the jet 
axis(\cite{pushkarev05, aaron99}) indicating a parallel magnetic field. However it should be 
noted here that the polarisation 
angle at the jet spine indicates a perpendicular magnetic field and this along with the parallel
magnetic field at the jet boundary can be an outcome of a dynamically dominant toroidal 
magnetic field structure
(\cite{pushkarev05,gabuzda99,gabuzda05}).
The radial velocity stratification of the jet can introduce Kelvin-Helmholtz instability and 
the stability of jets against this instability was studied by many authors(\cite{turland76, 
blandford76,ferrari78,hardee79,birkinshaw91}).

\cite{giroletti04} have studied the limb-brightened structure of
MKN501 jet considering the differential Doppler boosting at the jet spine and the boundary
(\cite{laing96,komissarov90}) and concluded the viewing angle (angle
between the jet and the line of sight of the observer) of the radio jet should be more than
$15^o$. However high energy studies of MKN501 demands the
viewing angle of the jet should be $\approx 5^o$ in order to explain the observed
rapid variability and the high energy emission(\cite{katar01, tavecchio01}). 
Considering the fact that the
gamma ray emission is originated from the inner part of the jet close to nucleus,
\cite{giroletti04} suggested a bending of the jet may
happen immediately after the gamma-ray zone to explain the required large viewing
angle of the radio jet. However the mechanism required to bend the jets are still
not well understood (jets deflected due to the pressure gradient in external medium 
is studied by \cite{canto96, raga96,mendoza01}) and  
moreover the observed large bending of the jet in the radio maps can be apparent one 
because of projection effects. This projection effects are even amplified
when the jet is close to the line of sight. Though it needs to be noted here that jets with large 
bending angle are indeed observed(\cite{savol}).

The limb-brightened structure can also be explained if we consider the synchrotron emission 
from the particles accelerated at the boundary and this inference does not require large 
viewing angle. \cite{eilek79,eilek82} considered the acceleration
of particles due to turbulence initiated by Kelvin-Helmholtz and Rayleigh-Taylor instabilities at
the jet boundary. Particles at the boundary can also be accelerated via shear acceleration 
(\cite{berezhko81a,berezhko81b}) and this case is considered in the present work. 
The acceleration of particles in a shear flow
or by turbulence is well studied by various authors for both relativistic and non relativistic case
(\cite{earl, webb89, ostrowski90, stawarz02, rieger06, stawarz08,virtanen05}).

In this letter we explain the observed limb-brightened feature of MKN501 by considering the 
diffusion of electrons accelerated at the jet boundary via shear acceleration. 
In the next section we show the required condition for the shear acceleration to be dominant over 
turbulent acceleration and in \S 3 we consider the diffusion of particles accelerated at the 
boundary into the jet medium. 
In \S 4 we discuss the spectral index of the particle distribution accelerated via shear acceleration 
process and turbulent acceleration process and show the observed index at the boundary of MKN501 jet
supports the earlier case.
Throughout this work, $H_o = 75$ km s$^{-1}$ Mpc$^{-1}$ 
and $q_0 = 0.5$ are adopted.

\section[]{Shear acceleration at MKN501 jet boundary}{\label{sec:shear}}
The particle acceleration process at the jet boundary can be described by the diffusion equation in 
momentum space. The evolution of an isotropic phase space distribution is given by(\cite{melrose68})
\begin{equation}
\label{eq:evol}
\frac{\partial f(p)}{\partial t}=\frac{1}{p^2}\frac{\partial}{\partial p} 
\left(p^2 D(p)\frac{\partial f(p)}{\partial p}\right)
\end{equation}
where $D(p)$ is the momentum diffusion coefficient. The characteristic acceleration timescale can 
be written as 
\begin{equation}
t_{acc}=p^3\left[\frac{\partial}{\partial p}\left( p^2 D(p)\right)\right]^{-1}
\end{equation}
If we consider a sheared flow, the electrons are scattered across different velocity layers by 
turbulent structures which are embedded in the shear flow. \cite{berezhko81a} showed in such case
there will be a net gain of energy in the electrons getting scattered and this process is referred 
as shear acceleration. The momentum diffusion coefficient in case of a shear flow
can be written as(\cite{rieger06,rieger07})
\begin{equation}
\label{eq:shdiffcoeff}
D_s(p) = \chi p^2 \tau
\end{equation}
where $\tau$ is the mean scattering time given by $\tau \simeq \lambda/c$ with $\lambda$ the mean free
path and $\chi$ is the shear coefficient given for a relativistic flow as(\cite{rieger04}) 
\begin{equation}
\chi = \frac{c^2}{15(\Gamma(r)^2-1)}\left(\frac{\partial \Gamma}{\partial r}\right)^2
\end{equation}
where $\Gamma(r)$ is the bulk Lorentz factor of the flow and $r$ is the radial coordinate of the jet
cross section. Using (\ref{eq:shdiffcoeff}), the shear acceleration timescale($t_{acc}^{(s)}$) 
for $\tau = \tau_o p^\xi$ will be
\begin{equation}
\label{eq:shacc}
t_{acc}^{(s)}= \frac{1}{(4+\xi)\chi \tau}
\end{equation}
In case of turbulent acceleration(stochastic), the particles are scattered off by randomly moving 
scattering centres and gets energized by second order Fermi acceleration. The momentum diffusion coefficient in this case 
can be approximated as(\cite{rieger07})
\begin{equation}
\label{eq:turbdiffcoeff}
D_t(p)\simeq \frac{p^2}{3\tau}\left(\frac{V_A}{c}\right)^2 
\end{equation}
where the Alfven velocity($V_A$) is given by
\begin{equation} 
V_A=\frac{B}{\sqrt{4\pi \rho}}
\end{equation}
here $B$ is the magnetic field and $\rho$ the mass density of the jet. 
Hence the turbulent acceleration timescale($t_{acc}^{(t)}$) 
will be
\begin{equation}
t_{acc}^{(t)}=\frac{3\tau}{(4-\xi)}\left(\frac{c}{V_A}\right)^2
\end{equation}
For shear acceleration to be dominant over turbulent acceleration $t_{acc}^{(s)}<t_{acc}^{(t)}$.
If we consider Bohm diffusion ($\xi = 1$) then the mean free path of the electron aligned 
to the magnetic field 
($\lambda_\parallel$) scales as the gyro radius ($r_g$)(\cite{abraham94}), 
$\lambda_\parallel \simeq \eta \frac{\gamma m_e c^2}{eB}$, where $\eta$ is a numerical factor
($\eta>1$ for magnetized particles) and
$\gamma(\gg 1)$
is the Lorentz factor of the electron scattered. Since the magnetic field at the jet boundary of 
MKN501 is parallel to the jet axis (or toroidal)(\cite{aaron99,pushkarev05,gabuzda99}), we
consider $\tau\simeq\lambda_\parallel/c$. Also if we consider
\begin{equation}
\frac{\partial \Gamma}{\partial r}\simeq \frac{\Delta \Gamma}{\Delta r}
\end{equation}
where $\Delta \Gamma$ is the difference between the bulk Lorentz factor at the jet spine and the 
jet boundary and $\Delta r$ is the thickness of the shear layer, then the condition for shear 
acceleration to be dominant over turbulent acceleration will be
\begin{equation}
\label{eq:shthick1}
\Delta r < \frac{\eta \gamma m_ec^3 (\Delta\Gamma)}{eB^2}\left[\frac{4\pi \rho}{3(\Gamma(r)^2-1)}\right]^
{\frac{1}{2}}
\end{equation} 
If we consider the mass density of the jet is dominated by cold protons
and if the number of protons are equal to the number of non-thermal electrons, then the 
jet mass density can be written in terms of equipartition magnetic field($B_{eq}$) as
\begin{equation}
\rho \simeq \frac{m_pB_{eq}^2(2\alpha-1)}{16 \pi m_ec^2 \alpha \gamma_{min}}
\end{equation}
and (\ref{eq:shthick1}) will be 
\begin{equation}
\label{eq:shthick2}
\Delta r < 0.29\frac{\eta \gamma c^2 (\Delta\Gamma)}{eB_{eq}}\left[\frac{m_em_p(2\alpha-1)}
{\alpha\gamma_{min}(\Gamma(r)^2-1)}\right]^{\frac{1}{2}}
\end{equation} 
where $\alpha$ is the observed photon spectral index, $m_p$ is the proton mass and $\gamma_{min}$ is
the Lorentz factor of electron responsible for the minimum observed 
photon frequency $\nu_{min}$. 
The equipartition magnetic field can be expressed in terms of observed 
quantities as 
\begin{equation}
B_{eq}\simeq 9.62\frac{1}{\Gamma(r)}(m_ece\nu_{min})^{\frac{1}{7}}\left[\frac{d_L^2F(\nu_{min})}
{V \sigma_T (2\alpha-1)}\right]^{\frac{2}{7}} G
\end{equation}
where $F(\nu_{min})$ is the flux at the minimum observed frequency $\nu_{min}$, $d_L$ is the
luminosity distance, $V$ is the volume of the emission region and $\sigma_T$ is Thomson cross section. 
Hence, for $\Gamma(r)^2\gg1$ and $\alpha\simeq 0.7$, shear acceleration will dominate the particle
spectrum at the jet boundary of MKN501 if the thickness of the shear layer
\begin{align}
\label{eq:shthick3}
\Delta r &< 7.22\times10^{-9}\times\left(\frac{\eta}{10}\right)
\left(\frac{\Delta \Gamma}{10}\right)
\left(\frac{\nu_{obs}}{1.6\textrm{GHz}}\right)^{\frac{1}{2}} 
\left(\frac{\nu_{min}}{10\textrm{MHz}}\right)^{-\frac{6}{14}} \times \nonumber \\
&\times
\left(\frac{F(10\textrm{MHz})}{910\textrm{mJy}}\right)^{\frac{-5}{14}}
\left(\frac{R}{1.5\textrm{parsec}}\right)^{\frac{15}{14}}
\textrm{parsec}
\end{align}
Where $R$ is the radius of the spherical region considered. (We assume $10 MHz$ as minimum 
observed frequency and the flux at $10 MHz$ is obtained from the flux at $1.6 GHz$ considering 
the same spectral index.  The flux at $1.6 GHz$ and $R$ in (\ref{eq:shthick3}) 
are obtained from a region around R.A $10\;mas$ and 
declination $-10\;mas$ from Fig.$7$ of \cite{giroletti04}). 
The corresponding equipartition 
magnetic field $B_{eq}$ for $\Gamma=5$ is $1.2\times 10^{-3} G$.

The electrons accelerated by shear acceleration cool via synchrotron radiation. The 
cooling time for synchrotron loss is given by
\begin{equation}
\label{eq:syncool}
t_{cool}=\frac{6\pi m_ec}{\gamma \sigma_T B_{eq}^2}
\end{equation}
Using (\ref{eq:shacc}) and (\ref{eq:syncool}), we find
\begin{align}
\label{eq:acccool}
\frac{t_{acc}^{(s)}}{t_{cool}}&\simeq 1.5 \times 10^{-12}\left(\frac{B}{1.2\times 10^{-3} G}\right)^3
\left(\frac{\eta}{10}\right)^{-1}\left(\frac{\Gamma(r)}{5}\right)^2 \times \nonumber \\
&\times \left(\frac{\Delta r}{10^{-9}parsec}\right)^2\left(\frac{\Delta \Gamma}{10}\right)^{-2}
\end{align}
and since $t_{acc}^{(s)}\ll t_{cool}$, shear acceleration dominates over synchrotron cooling. 
It can be noted that 
(\ref{eq:acccool}) is independent of the electron energy and hence the maximum energy of the electron
will be decided by the loss processes other than synchrotron loss (which are not considered in 
this simplistic treatment).

If we maintain the general form of mean scattering time $\tau=\tau_0p^\xi$, then for shear 
acceleration to dominate over turbulent acceleration the thickness of the shear layer 
($\Delta r$) should be
\begin{equation}
\label{eq:shthickgen}
\Delta r < 1.7\times 10^6 \frac{\tau_0 p^\xi (\Delta \Gamma)}{\Gamma(r)}
\left[\frac{(4+\xi)(2\alpha -1)}{\alpha (4-\xi) \gamma_{min}}\right]^{\frac{1}{2}}
\textrm{cm}
\end{equation}
It can be noted that (\ref{eq:shthick1}) is equal to (\ref{eq:shthickgen}) if we set in the latter
$\xi=1$ and $\tau_0 p^\xi = \eta r_g/c$.  

\section[]{Particle Diffusion at the jet boundary and Limb-brightening}

Particles accelerated at the shear layer of the jet boundary, diffuse into the jet medium before
getting cooled off via synchrotron radiation. As the magnetic field at the jet boundary is 
parallel to the jet axis (or toroidal)(\cite{aaron99,pushkarev05,gabuzda99}), the radial diffusion 
of the electron into the jet medium is determined by cross field diffusion.  
The cross field diffusion coefficient can be approximated as (\cite{axford65, jokipii87, abraham94})  
\begin{equation}
\label{eq:crossdiff}
\kappa_\perp \approx \frac{1}{3\eta} r_g c
\end{equation}
Where $\eta(>1)$ is the scaling factor determining the field aligned mean free path 
(see (\S\ref{sec:shear})).

The radial distance $R_{diff}$ that the electron diffuse before getting cooled can then be 
approximated as 
\begin{equation}
\label{eq:rdiff0}
R_{diff} \approx \sqrt{\kappa_\perp t_{cool}}
\end{equation}
Using (\ref{eq:syncool}) and (\ref{eq:crossdiff}) and considering the equipartition magnetic 
field we get
\begin{equation}
\label{eq:rdiff}
R_{diff} \simeq 2.9 \times 10^{-4} \left(\frac{\eta}{10}\right)^{-\frac{1}{2}} 
\left(\frac{B}{1.2\times 10^{-3} G}\right)^{-\frac{3}{2}} \textrm{parsec}
\end{equation}

Since the thickness of the shear layer $\Delta r\ll R_{diff}$ (refer (\ref{eq:shthick3}) and 
(\ref{eq:rdiff})), the thickness of the limb brightened structure will be $\approx R_{diff}$. 
This corresponds to an angular distance of $4.7\times10^{-4}mas$ which is beyond the resolution
of present day telescopes.

For $\tau=\tau_0p^\xi$, the cross field diffusion coefficient will be 
\begin{equation}
\kappa_\perp\simeq\frac{1}{3\tau_0}r_g^2 p^{-\xi}
\end{equation}
Using (\ref{eq:syncool}) and (\ref{eq:rdiff0}) we get 
\begin{equation}
R_{diff} \simeq 5.2 \times 10^{15} B^{-2}\tau_0^{-\frac{1}{2}} p^{\frac{1-\xi}{2}}\textrm{cm}
\end{equation}
and hence the thickness of the limb brightened structure will be energy dependent for $\xi \ne 1$.

\section[]{Spectral index}{\label{sec:spectral}}
If we add mono-energetic particle injection term ($\delta(p-p_o)$) and particle 
escape term ($-1/t_{esc}$)
in (\ref{eq:evol}), then the steady state equation in case of shear acceleration 
for $p>p_o$ and  $\xi=1$ can be written as 
\begin{equation}
\label{eq:shearsteady1}
p^3\frac{d^2f_s}{dp^2}+5p^2\frac{df_s}{dp}-\frac{f_s}{\chi\tau_o t_{esc}} = 0
\end{equation}
and in case of turbulent acceleration it will be  
\begin{equation}
\label{eq:turbsteady}
p\frac{d^2f_t}{dp^2}+3\frac{df_t}{dp}-\frac{f_t}{\psi t_{esc}} = 0
\end{equation}
where $\psi = \frac{V_A^2}{3c^2\tau_o}$. If we substitute $p = 1/x$ in (\ref{eq:shearsteady1})
we get 
\begin{equation}
\label{eq:shearsteady}
x\frac{d^2f_s}{dx^2}-3\frac{df_s}{dx}-\frac{f_s}{\chi \tau_o t_{esc}} = 0
\end{equation}
Equations (\ref{eq:turbsteady}) and (\ref{eq:shearsteady}) can be solved analytically 
(\cite{kepinski}) and the solutions are complex and are given by
\begin{align}
\label{eq:shearsol}
f_s&=\left(\frac{1}{\chi\tau_o p t_{esc}}\right)^2 \times \nonumber \\
	 &\times \left[a_s J_4\left(2i\sqrt{\frac{1}
	{\chi\tau_o p t_{esc}}}\right)+b_s Y_4\left(2i\sqrt{\frac{1}{\chi\tau_o p t_{esc}}}\right)\right]
\end{align}
and
\begin{equation}
\label{eq:turbsol}
f_t=\left(\frac{\psi t_{esc}}{p}\right)\left[a_t J_2\left(2i\sqrt{\frac{p}{\psi t_{esc}}}
\right)+b_t Y_2\left(2i\sqrt{\frac{p}{\psi t_{esc}}}\right)\right]
\end{equation}
Where $J_n(z)$ and $Y_n(z)$ are the Bessel functions of first and second kind and $a_s$,
$b_s$, $a_t$ and $b_t$ are constants. For negligible escape ($t_{esc} \to \infty$), using the limiting
forms of Bessel functions (\cite{abramovitz}), the solutions (\ref{eq:shearsol}) and (\ref{eq:turbsol})
approaches a power law $f_s \propto p^{-4}$ and $f_t \propto p^{-2}$. The shear accelerated 
particle number density will then be $n_s(p)\propto p^{-2}$ and the corresponding synchrotron photon
flux will be $S_{\nu,shear} \propto \nu ^{-1/2}$. For turbulent acceleration the number density will be 
independent of $p$ ($n_t(p)\propto p^0$) and hence the observed synchrotron photon flux will be 
a flat one $S_{\nu,turb} \propto \nu ^{1/3}$.
The spectral index map of MKN501 jet indicates a steep photon spectra at the boundary 
and a flat spectra at the spine(\cite{giroletti04}). Hence it can be argued that the 
shear acceleration may be dominant
at the jet boundary of MKN501 and turbulent acceleration at the jet spine. However $\xi$ is
usually related to the turbulent spectral index (\cite{biermann87}) which may be different at the
jet boundary and jet spine.
  
\section[]{Discussion}
As the AGN jet moves through the ambient medium the viscosity involved will cause a shear at the jet
boundary and hence acceleration of particles in these shear layer is unavoidable. If the 
shear gradient $\partial \Gamma/\partial r$ is very steep or if the shear layer is very thin
(\ref{eq:shthick3}), then shear acceleration can dominate over the turbulent acceleration initiated 
by the instabilities at the jet boundary(\cite{eilek82}). Turbulent acceleration may play
an important role at the interior regions of the jet (\cite{virtanen05}) and can provide an alternative to 
explain the emission from the inter knot regions of AGN jets (\cite{macchetto96, jester01}). 
The observed hard spectra at the jet spine (\cite{giroletti04}) also supports this inference since 
turbulent acceleration can produce a hard particle spectra(\cite{virtanen05}) (also shown in section
\S(\ref{sec:spectral}).
The electrons accelerated by the turbulence can be reaccelerated by shocks and can form a broken 
power law electron spectrum. This can possibly
explain the break in the radio-to-x-ray spectra of the knots of FRI jets (\cite{sahayanathan08}). 

\cite{giroletti04} calculated the jet viewing angle ($\theta$) using the correlation between the core 
power and the total power (\cite{giovanni01}). They estimated the jet viewing angle to be within 
$10^o<\theta<27^o$ by comparing the observed core radio power and the expected intrinsic core power 
derived from the correlation. However this estimation may vary if the core flux density variability 
is more than factor $2$. Also considering the variation of the parameter values in the correlation with
increase number of samples, this may not provide a strong constrain on the jet viewing angle. The 
estimate of $\theta$ based on the adiabatically expanding relativistic jet model (\cite{baum97}) may 
not be a strong constraint as it considers a simplified situation. Also the constrain is less severe in case
of perpendicular magnetic fields and observed polarisation studies have indicated the presence of
perpendicular magnetic fields at jet spine (\cite{pushkarev05,aaron99}). \cite{stawarz02} proposed a 
model similar to the present one, however their aim was to show the observational implications of the 
two-component particle spectrum (power law distribution with high energy pile-up) formed at the
boundary shear layer and the complex beaming pattern.

\section[]{Conclusion}
The observed limb brightened structure seen in the radio maps
of MKN501 jet can be explained if we consider the shear acceleration of particles at the 
boundary due to velocity stratification and their diffusion into the jet medium. This 
inference does not demand large viewing angle which is  
required otherwise for the explanation via differential Doppler boosting of the jet spine and boundary.
We have shown that shear acceleration dominates over turbulent acceleration
at the boundary if we consider thin shear layer or a sharp velocity gradient. Also for the 
estimated set of parameters, shear acceleration timescale is much smaller than synchrotron cooling
timescale allowing acceleration of electrons to be possible. The thickness
of the limb brightened structure will be decided by the distance electrons have diffused into 
the jet medium before loosing its energy via synchrotron radiation. However the estimated thickness 
is beyond the resolution of present day telescopes. Simple analytical solution
of the steady state diffusion equation considering mono-energetic injection and  
particle escape, indicates a steep particle spectra for the electrons accelerated at the shear 
layer in comparison with turbulent acceleration. The radio spectral index map of MKN501 jet is
also observed to have steep spectra at the boundary supporting the presence of shear acceleration. 

The author is grateful to the anonymous referee for his useful comments which helped in clearing many of
the ignorances and a better understanding. The author acknowledges the useful discussions with 
S. Bhattacharyya, N. Bhatt, M.Choudhury  and A.Mitra. The author is grateful to 
L. Stawarz and F. M. Rieger for enlightening information on various topics related to shear acceleration.

\label{lastpage}


\begin{thebibliography}{99}

\bibitem[\protect\citeauthoryear{Aaron}{1999}]{aaron99} 
  Aaron S., 1999, in Takalo L. O., Sillanpaa A., eds., ASP Conf. Ser. Vol. 159, BL Lac Phenomenon.
 Astron. Soc. Pac., San Francisco, p. 427 

\bibitem[\protect\citeauthoryear{Abramowitz 
\& Stegun}{1972}]{abramovitz} Abramowitz M., Stegun I. A., 1972, Handbook of Mathematical Functions. 
Dover Publications, New York  

\bibitem[\protect\citeauthoryear{Achterberg 
\& Ball}{1994}]{abraham94} Achterberg A., Ball L., 1994, A\&A, 285, 687 

\bibitem[\protect\citeauthoryear{Aloy et al.}{2000}]{aloy00} 
Aloy M.-A., G{\'o}mez J.-L., Ib{\'a}{\~n}ez J.-M., Mart{\'{\i}} J.-M., M{\"u}ller E., 2000, ApJ, 528, L85 

\bibitem[\protect\citeauthoryear{Axford}{1965}]{axford65} 
  Axford W. I., 1965, Planetary Space Sci., 13, 115 

\bibitem[\protect\citeauthoryear{Baum et al.}{1997}]{baum97} 
Baum S.~A., et al., 1997, ApJ, 483, 178 

\bibitem[\protect\citeauthoryear{Berezhko}{1981}]{berezhko81a} 
Berezhko E.~G., 1981, J. Exp. Tjeor. Phys. Lett., 33, 399 

\bibitem[\protect\citeauthoryear{Berezhko 
\& Krymskii}{1981}]{berezhko81b} Berezhko E.~G., Krymskii G.~F., 1981, SvA, 7, L352 

\bibitem[\protect\citeauthoryear{Biermann 
\& Strittmatter}{1987}]{biermann87} Biermann P.~L., Strittmatter P.~A., 1987, ApJ, 322, 643 

\bibitem[\protect\citeauthoryear{Birkinshaw}{1991}]{birkinshaw91} 
Birkinshaw M., 1991, MNRAS, 252, 505 

\bibitem[\protect\citeauthoryear{Blandford 
\& Pringle}{1976}]{blandford76} Blandford R.~D., Pringle J.~E., 1976, MNRAS, 176, 443 

\bibitem[\protect\citeauthoryear{Bloom 
\& Marscher}{1996}]{bloom96} Bloom S.~D., Marscher A.~P., 1996, ApJ, 461, 657 

\bibitem[\protect\citeauthoryear{B{\"o}ttcher}{2000}]{bottcher00} 
B{\"o}ttcher M., 2000, in Dingus B. L., Salamon M. H., Kieda D. B., eds., Proc. AIP Conf. Vol. 515, 
GeV-TeV GAMMA RAY ASTROPHYSICS WORKSHOP: Towards a Major Atmospheric Cherenkov Detector VI. 
AIP Conf. Proc., Snowbird, UT, p. 31 

\bibitem[\protect\citeauthoryear{Canto 
\& Raga}{1996}]{canto96} Canto J., Raga A.~C., 1996, MNRAS, 280, 559 

\bibitem[\protect\citeauthoryear{Coppi 
\& Aharonian}{1999}]{coppi99} Coppi P.~S., Aharonian F.~A., 1999, ApJ, 521, L33

\bibitem[\protect\citeauthoryear{Costamante 
\& Ghisellini}{2002}]{costa02} Costamante L., Ghisellini G., 2002, A\&A, 384, 56 

\bibitem[\protect\citeauthoryear{Dondi 
\& Ghisellini}{1995}]{dondi95} Dondi L., Ghisellini G., 1995, MNRAS, 273, 583 

\bibitem[\protect\citeauthoryear{Earl, Jokipii, 
\& Morfill}{1988}]{earl} Earl J.~A., Jokipii J.~R., Morfill G., 1988, ApJ, 331, L91 

\bibitem[\protect\citeauthoryear{Edwards et al.}{2000}]{edward00} Edwards P.~G., Giovannini G., Cotton 
W.~D., Feretti L., Fujisawa K., Hirabayashi H., Lara L., Venturi T., 2000, PASJ, 52, 1015 

\bibitem[\protect\citeauthoryear{Eilek}{1979}]{eilek79} Eilek 
J.~A., 1979, ApJ, 230, 373 

\bibitem[\protect\citeauthoryear{Eilek}{1982}]{eilek82} Eilek 
J.~A., 1982, ApJ, 254, 472 

\bibitem[\protect\citeauthoryear{Ferrari, Trussoni, 
\& Zaninetti}{1978}]{ferrari78} Ferrari A., Trussoni E., Zaninetti L., 1978, A\&A, 64, 43 

\bibitem[\protect\citeauthoryear{Gabuzda, Murray, 
\& Cronin}{2005}]{gabuzda05} Gabuzda D.~C., Murray {\'E}., Cronin P., 2005, BaltA, 14, 363 

\bibitem[\protect\citeauthoryear{Gabuzda}{1999}]{gabuzda99} 
Gabuzda D.~C., 1999, NewAR, 43, 691 

\bibitem[\protect\citeauthoryear{Gaidos et al.}{1996}]{gaidos96} 
Gaidos J.~A., et al., 1996, Natur, 383, 319 

\bibitem[\protect\citeauthoryear{Ghisellini et al.}{1993}]{ghisellini93} 
Ghisellini G., Padovani P., Celotti A., Maraschi L., 1993, ApJ, 407, 65

\bibitem[\protect\citeauthoryear{Giovannini et al.}{2001}]{giovanni01} 
Giovannini G., Cotton W.~D., Feretti L., Lara L., Venturi T., 2001, ApJ, 552, 508 

\bibitem[\protect\citeauthoryear{Giovannini et al.}{1999}]{giovanni99} 
Giovannini G. et al., 1999, in Takalo L. O., Sillanpaa A., eds, ASP Conf.
Ser. Vol. 159, BL Lac Phenomenon. Astron. Soc. Pac., San Francisco,
p. 439

\bibitem[\protect\citeauthoryear{Giroletti et 
al.}{2004}]{giroletti04} Giroletti M., et al., 2004, ApJ, 600, 127 

\bibitem[\protect\citeauthoryear{Hardee}{1979}]{hardee79} Hardee 
P.~E., 1979, ApJ, 234, 47 

\bibitem[\protect\citeauthoryear{Jester et al.}{2001}]{jester01} 
Jester S., R{\"o}ser H.-J., Meisenheimer K., Perley R., Conway R., 2001, A\&A, 373, 447

\bibitem[\protect\citeauthoryear{Jokipii}{1987}]{jokipii87} 
Jokipii J.~R., 1987, ApJ, 313, 842 

\bibitem[\protect\citeauthoryear{Kahn}{1983}]{kahn83} Kahn 
F.~D., 1983, MNRAS, 202, 553 

\bibitem[\protect\citeauthoryear{Kataoka et 
al.}{1999}]{katoka99} Kataoka J., et al., 1999, ApJ, 514, 138 

\bibitem[\protect\citeauthoryear{Katarzy{\'n}ski, Sol, 
\& Kus}{2001}]{katar01} Katarzy{\'n}ski K., Sol H., Kus A., 2001, A\&A, 367, 809

\bibitem [\protect\citeauthoryear{Kepinski}{1905}]{kepinski} 
  Kepinski, S., 1905, Math. Ann. (Springer), 61, 397

\bibitem[\protect\citeauthoryear{Komissarov}{1990}]{komissarov90} 
Komissarov S.~S., 1990, SvA, 16, L284 

\bibitem[\protect\citeauthoryear{Krawczynski}{2004}]{kraw04} 
Krawczynski H., 2004, NewAR, 48, 367 

\bibitem[\protect\citeauthoryear{Krawczynski et 
al.}{2000}]{kraw00} Krawczynski H., Coppi P.~S., Maccarone T., Aharonian F.~A., 2000, A\&A, 353, 97 

\bibitem[\protect\citeauthoryear{Laing}{1996}]{laing96} 
Laing R. A., 1996, in Hardee P. E., Bridle A. H., Zensus J. A., eds, ASP
Conf. Ser. Vol. 100, Energy Transport in Radio Galaxies and Quasars.
Astron. Soc. Pac., San Francisco, p. 241  

\bibitem[\protect\citeauthoryear{Leismann et al.}{2005}]{leismann05} 
Leismann T., Ant{\'o}n L., Aloy M.~A., M{\"u}ller E., Mart{\'{\i}} J.~M., 
Miralles J.~A., Ib{\'a}{\~n}ez J.~M., 2005, A\&A, 436, 503 

\bibitem[\protect\citeauthoryear{Macchetto}{1996}]{macchetto96} 
 Macchetto F. D., 1996, in Ekers R. D., Fanti C., Padrielli L., eds, Proc. IAU
Symp. 175, Extragalactic Radio Sources. Kluwer, Dordrecht, p. 195

\bibitem[\protect\citeauthoryear{Mannheim}{1998}]{mannheim98} 
Mannheim K., 1998, Sci, 279, 684 

\bibitem[\protect\citeauthoryear{Maraschi, Ghisellini, 
\& Celotti}{1992}]{maraschi92} Maraschi L., Ghisellini G., Celotti A., 1992, ApJ, 397, L5

\bibitem[\protect\citeauthoryear{Melrose}{1968}]{melrose68} Melrose D.~B., 1968, Ap\&SS, 2, 171 

\bibitem[\protect\citeauthoryear{Mendoza 
\& Longair}{2001}]{mendoza01} Mendoza S., Longair M.~S., 2001, MNRAS, 324, 149 

\bibitem[\protect\citeauthoryear{M{\"u}cke et al.}{2003}]{mucke03} 
M{\"u}cke A., Protheroe R.~J., Engel R., Rachen J.~P., Stanev T., 2003, APh, 18, 593 

\bibitem[\protect\citeauthoryear{Ostrowski}{1990}]{ostrowski90} Ostrowski M., 1990, A\&A, 238, 435 

\bibitem[\protect\citeauthoryear{Owen, Hardee, 
\& Cornwell}{1989}]{owen89} Owen F.~N., Hardee P.~E., Cornwell T.~J., 1989, ApJ, 340, 698 

\bibitem[\protect\citeauthoryear{Perlman et al.}{1999}]{perlman99} 
Perlman E.~S., Biretta J.~A., Zhou F., Sparks W.~B., Macchetto F.~D., 1999, AJ, 117, 2185 

\bibitem[\protect\citeauthoryear{Pushkarev et al.}{2005}]{pushkarev05} 
Pushkarev A.~B., Gabuzda D.~C., Vetukhnovskaya Y.~N., Yakimov V.~E., 2005, MNRAS, 356, 859 

\bibitem[\protect\citeauthoryear{Quinn et al.}{1996}]{quinn96} 
Quinn J., et al., 1996, ApJ, 456, L83 

\bibitem[\protect\citeauthoryear{Raga 
\& Canto}{1996}]{raga96} Raga A.~C., Canto J., 1996, MNRAS, 280, 567 

\bibitem[\protect\citeauthoryear{Rieger 
\& Duffy}{2004}]{rieger04} Rieger F.~M., Duffy P., 2004, ApJ, 617, 155 

\bibitem[\protect\citeauthoryear{Rieger 
\& Duffy}{2006}]{rieger06} Rieger F.~M., Duffy P., 2006, ApJ, 652, 1044 

\bibitem[\protect\citeauthoryear{Rieger, Bosch-Ramon, 
\& Duffy}{2007}]{rieger07} Rieger F.~M., Bosch-Ramon V., Duffy P., 2007, Ap\&SS, 309, 119 

\bibitem[\protect\citeauthoryear{Sambruna}{2000}]{samb00} 
Sambruna R. M., 2000, in Dingus B. L., Salamon M. H., Kieda D. B., eds,
Proc. AIP Conf. 515, GeV-TeV Gamma Ray Astrophysics Workshop:
Towards a Major Atmospheric Cherenkov Detector VI. Am. Inst. Phys.,
New York, p. 19  

\bibitem[\protect\citeauthoryear{Savolainen et 
al.}{2006}]{savol} Savolainen T., Wiik K., Valtaoja E., 
Kadler M., Ros E., Tornikoski M., Aller M.~F., Aller H.~D., 2006, ApJ, 647, 
172 

\bibitem[\protect\citeauthoryear{Stawarz 
\& Ostrowski}{2002}]{stawarz02} Stawarz {\L}., Ostrowski M., 2002, ApJ, 578, 763 

\bibitem[\protect\citeauthoryear{Stawarz 
\& Petrosian}{2008}]{stawarz08} Stawarz {\L}., Petrosian V., 2008, ApJ, 681, 1725 

\bibitem[\protect\citeauthoryear{Sahayanathan}{2008}]{sahayanathan08} 
Sahayanathan S., 2008, MNRAS, 388, L49 

\bibitem[\protect\citeauthoryear{Tavecchio et 
al.}{2001}]{tavecchio01} Tavecchio F., et al., 2001, ApJ, 554, 725 

\bibitem[\protect\citeauthoryear{Turland 
\& Scheuer}{1976}]{turland76} Turland B.~D., Scheuer P.~A.~G., 1976, MNRAS, 176, 421 

\bibitem[\protect\citeauthoryear{Urry 
\& Padovani}{1995}]{urrypado95} Urry C.~M., Padovani P., 1995, PASP, 107, 803 

\bibitem[\protect\citeauthoryear{Virtanen 
\& Vainio}{2005}]{virtanen05} Virtanen J.~J.~P., Vainio R., 2005, ApJ, 621, 313 

\bibitem[\protect\citeauthoryear{Webb}{1989}]{webb89} Webb 
G.~M., 1989, ApJ, 340, 1112 

\end{thebibliography}
\end{document}